%% file: Linearitycorr.tex
\documentclass[submission,copyright,creativecommons]{eptcs}

\input{0_commands}

\providecommand{\urlalt}[2]{\href{#1}{#2}}
\providecommand{\doi}[1]{doi:\urlalt{http://dx.doi.org/#1}{#1}}
\newcommand{\arxiv}[1]{\urlalt{http://arxiv.org/abs/#1}{arXiv:#1}}

\providecommand{\event}{Linearity 2016} 
\usepackage{breakurl}             

\title{Proof Diagrams for Multiplicative Linear Logic}
\author{Matteo Acclavio
\institute{I2M\\ Marseille, France}
\institute{Aix-Marseile Universit\'e}
\email{matteo.acclavio@univ-amu.fr}
}
\def\titlerunning{Proof Diagrams for Multiplicative Linear Logic}
\def\authorrunning{Matteo Acclavio}

\begin{document}
\maketitle

\begin{abstract}
The original idea of proof nets can be formulated by means of interaction nets syntax. Additional machinery as switching, jumps and graph connectivity is needed in order to ensure correspondence between a proof structure and a correct proof in sequent calculus.

In this paper we give an interpretation of proof nets in the syntax of string diagrams. Even though we lose standard proof equivalence, our construction allows to define a framework where soundness and well-typeness of a diagram can be verified in linear time.
\end{abstract}

\section*{Introduction}
Proof nets are a geometrical representation of \emph{linear logic} proofs  introduced by J-Y.Girard \cite{girll}. The building blocks of proof nets are called \emph{proof structures} that have been generalized by Y. Lafont~\cite{lafint} in the so-called \emph{interaction nets}. 
To recognize if a proof structure is a proof net one needs to verify its \emph{sequentializability} property, that is, whether it corresponds to a linear logic proof derivation.
Following Girard's original correction criterion, others methods have been introduced, notably by Danos-Regnier \cite{dan-reg}, that ensures graph acyclicity by a notion of  \emph{switchings} on $\otimes$ cells, and by Guerrini \cite{guerrini}, that reformulates correction by means of graph contractability.

Proof structures allow to recover the semantic equivalence of derivation under commutation and permutation of some inference rules. Unfortunately this property makes  ineffective the aforementioned criteria in presence of the multiplicative unit $\bot$. In order to recover a sequentialization condition for the multiplicative fragment with units, Girard has introduced the notion of \emph{jumps} \cite{girjump}. These are untyped edges between two cells  which express a \emph{dependency} relation of the respective rules in sequentialization.

In this work we reformulate the proof net idea of a $2$-dimensional representation of proofs by replacing the underlying interaction nets syntax with that of string diagrams in order to achieve a new sequentialization criterion. String diagrams \cite{baez} are a syntax for $2$-arrows (or $2$-cells) of a $2$-category with a rigid structure. Although the two syntaxes may graphically look similar, string diagrams' \emph{strings} do not just denote connections between cells but they represent morphisms.
Since crossing strings is not allowed without the introduction of \emph{twisting operators}, we introduce the notion of \emph{twisting relations} in order to equate diagrams by permitting cells to cross certain strings.

We study several diagram rewriting systems given by \emph{twisting polygraphs}, a particular class of polygraph \cite{Bur} where string crossings are restrained by the introduction of some non-crossing \emph{control strings} in the syntax.

As soon as one considers proof derivations as sequences of $n$-ary operators applications over lists of formulas, then control strings intuitively represent their correct parenthesization. In particular these strings disallow  non-sequentializable diagram compositions, lastly resulting, thanks to negative units' fixed position, into a sound framework where sequentializability depends on diagram inputs and outputs pattern only. 
Moreover, this model gives a categorical semantics for linear logic proofs different from the standard one (see \cite{Mel}).

\section{String diagrams}
%
%

%
%
%

\subsection{Monochrome String Diagrams}
We now recall some basic notions in string diagram rewriting by considering the \emph{monochrome string diagrams} settings, where there are no labels on backgrounds or strings. For an introduction to string diagrams, see J. Baez's notes \cite{baez}.

Given $p$ and $q$ natural numbers, a diagram $\phi : p \frr q$ with $p$ \emph{inputs} and $q$ \emph{outputs} is pictured as follows:
$$
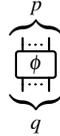
\overbrace{\underbrace{
\twocell{d *1 phi *1 d}}_{q}}^{p}$$
Diagrams may be composed in two different ways. If $\phi : p \frr q$ and $\phi': p' \frr q'$ are diagrams, we define:
\begin{itemize}

\item \emph{sequential} composition: if $q=p'$, the diagram $\phi' \circ \phi : p\frr q'$ corresponds to usual composition of maps.

This composition is associative with unit $\id_p: p\frr p$ for each $p\in \N$. In other words, we have $\phi \circ \id_p= \phi= \id_q \circ \phi$. The\emph{ identity diagram} $id_p$ is pictured as follows: $\underbrace{
\twocell{d}}_{p}$

\item \emph{parallel}  composition: the diagram $\phi * \phi': p+p' \frr q+q'$ is always defined.
This composition is associative with unit $\id_0: 0 \frr 0$. In other words, we have $\id_0 * \phi=\phi=\phi * \id_0$. This $\id_0$ is called the \emph{empty diagram}.
\end{itemize}

These two compositions are respectively represented as follows:
$$
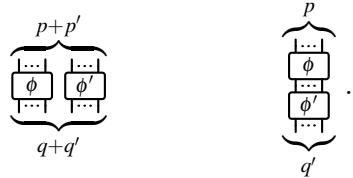
\overbrace{\underbrace{
\twocell{(d *0 d)*1( phi *0 phi1)*1(d *0 d)}}_{q+q'}}^{p+p'}
\qquad \qquad \qquad
\overbrace{\underbrace{
\twocell{d *1 phi *1 d *1 phi1 *1 d}}_{q'}}^{p}\; .
$$

Our two compositions satisfy the \emph{interchange rule}: if $\D:p\frr q$ and $\D' : p'\frr q'$, so $ (\id_q * \phi' ) \circ (\phi * \id_{p'})= \phi * \phi' =(\phi * \id_{q'})\circ (\id_p * \phi')$ that corresponds to the following picture:

$$\twocell{(d *0 d) *1 (phi *0 2) *1 (2 *0 phi1) *1 (d *0 d)}=\twocell{(d *0 d) *1 (phi *0 phi1) *1 (d *0 d)}=\twocell{(d *0 d) *1  (2 *0 phi1) *1 (phi *0 2) *1 (d *0 d)}$$

Monochrome string diagrams can be interpreted as morphisms in a $\mathbf{PRO}$, that is a strict monoidal category whose objects are natural numbers and whose product  on objects is addition. To be coherent with the cellular notation we use in next sections, diagrams represent 2-arrows in the $\mathbf{2-PRO}$ obtained by suspension of a regular  $\mathbf{PRO}$ (see \cite{GuirMalHD}).

\begin{defi}[Signature]
A \emph{signature} $\sig$ is a finite set of \emph{atomic diagrams} (or \emph{gates type}). Given a signature, a diagram $\phi : p \frr q$ is a morphism in the  $\mathbf{PRO}$ $\sig ^*$ freely generated by $\sig$, i.e. by the two compositions and identities. A \emph{gate} is an occurrence of an atomic diagram, we note $g: \alpha$ if $g$ is an occurrence of $\alpha \in \sig$.

\end{defi}

\begin{defi}
We say that $\phi$  is a \emph{subdiagram} of $\phi'$ whenever there exist $ \psi_u,\psi_d\in \sig^*$ and $k,k' \in \N$ such that $\phi'=\psi_d \circ (\id_\Gamma *\phi*\id_{\Delta})\circ \psi_u $.
\end{defi}

\nota Given $\phi\in \sig^*$ and $\sig' \subseteq \sig$, we write $|\phi|_{\sig'}$ the number of gates in $\phi$ with gate type $\alpha \in \sig'$.

\begin{defi}
We call \emph{horizontal} a diagram $\phi$ generated by parallel composition (and identities) only in $\sig^*$. It is \emph{elementary} if $|\phi|_\sig=1$.
\end{defi}

\subsection{Diagram rewriting}
\begin{defi}[Diagram Rewriting System]
A \emph{diagram rewriting system} is  a couple $(\sig, \R)$ given by a signature $\sig$ and a set $\R$ of rewriting rules of the form
$$
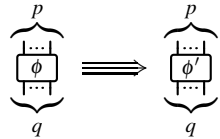
\overbrace{\underbrace{
\twocell{d *1 phi *1 d}}_q}^p
\xymatrix{\ar@3[r]&}
\overbrace{\underbrace{
\twocell{d *1 phi1 *1 d}}_q}^p
$$
where $\D, \D': p \frr q $ are diagrams in $\sig^*$. 
\end{defi}

\begin{defi}
We allow each rewriting  rules under any context, that is, if  $\xymatrix@C=1em{ \phi \ar@3[r] & \phi'}$ in $\R$ then, for every $\chi_u , \chi_d\in \sig^*$,
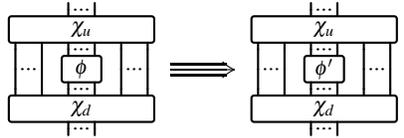
$$
\twocell{d *1 up *1 (2 *0 d *0 2) *1 (d *0 phi *0 d ) *1 ( 2 *0 d *0 2) *1 do *1 d}
\xymatrix{\ar@3[r]&}
\twocell{d *1 up *1 (2 *0 d *0 2) *1 (d *0 phi1 *0 d ) *1 ( 2 *0 d *0 2) *1 do *1 d} \; .$$
 We say that $\psi$ \emph{reduces}, or \emph{rewrites}, to $\psi'$  (denoted $\xymatrix@C=1em{\psi \ar@3[r]^{*} & \psi'}$) if there is a \emph{rewriting sequence} $P: \xymatrix@C=1em{\psi=\psi_0 \ar@3[r] & \psi_1 \ar@3[r] & \dots  \ar@3[r] & \psi_n=\psi'}$.
\end{defi}

We here recall some classical notions in rewriting:

\begin{itemize}
\item A diagram $\phi$ is \emph{irreducible} if there is no $\phi'$ such that $\xymatrix@C=1em{\phi \ar@3[r] & \phi'}$;

\item A rewriting system \emph{terminates} if there is  no infinite rewriting sequence;

\item A rewriting system is \emph{confluent} if for all $\phi_1,\phi_2$ and $\D$ such that $\xymatrix@C=1em{\phi \ar@3[r] & \phi_1}$ and $\xymatrix@C=1em{\phi \ar@3[r] & \phi_2 }$, there exists $\phi' $ such that $\xymatrix@C=1em{\phi_1 \ar@3[r]^{*} & \phi'} $  and $\xymatrix@C=1em{\phi_2 \ar@3[r]^{*} & \phi'}$;

\item A rewriting system is \emph{convergent} if both properties hold.

\end{itemize}
%

\section{Polygraphs}

In this section we formulate some basic notion by using the language of \emph{polygraphs}. 
Introduced by Street \cite{Street} as \emph{computads}, later reformulated and extended by Burroni \cite{Bur}, polygraphs can be considered as the generalization, for higher dimensional categories, of the notion of monoid presentation. 

Here we study some diagram rewriting systems with labels on strings in terms of $3$-polygraphs, which are denoted $\s=(\s_0,\s_1,\s_2,\s_3)$.
In particular, we consider polygraphs with just one $0$-cell in $\s_0$ in order to avoid background labeling. The set of $1$-cells $\s_1$ represents string labels, the $2$-cells in $\s_2$ are the  signature $\sig_\s$ of our rewriting system with rules $\R_\s=\s_3$, the set of $3$-cells. We say that a polygraph $\s$ exhibits some computational properties when the relative diagram rewriting system does. 

\nota   We denote $\phi\in \s$ whenever $\phi $ is a diagram generated by the associated signature $\sig_\s$.

\subsection{Twisting Polygraph}
In this section we introduce a notion of polygraph which generalizes polygraphic presentations of symmetric monoidal categories.

\begin{defi}[Symmetric polygraph]
We call the \emph{polygraph of permutation} the following monochrome $3$-polygraph:
$\mathfrak S=\biggl(\s_0=\{\square\}, \s_1=\{\twocell{1}\} , \s_2=\{\swap \}, \s_3=\biggl\{ \xymatrix@C-1pc{ \twocell{s *1 s } \ar@3[r] & \twocell{2}}, \; 
\xymatrix@C-1pc{ {\twocell{(s *0 1)*1 (1 *0 s)*1 (s *0 1)}} \ar@3[r] & {\twocell{ (1 *0 s)*1 (s *0 1)*1 (1 *0 s) }}}\biggr\} \biggr)$.

We call \emph{symmetric} a  $3$-polygraph $\s$ with one $0$-cell, one $1$-cell (i.e. $\s_1=\{ \twocell {1}\}$), containing one $2$-cell $\twocell{s} \in \s_2$ and such that the following holds
\vspace{-.3cm}
$$\twocell{s *1 s } = \twocell{2} \quad \mbox{, }\quad \twocell{(d *0 1) *1 (a *0 1) *1 (d*0 1) *1 rn *1 (1*0 d)} =  \twocell{ (d*0 1) *1 rn *1 (1*0 d) *1 (1*0 a)*1 (1 *0 d)}
 \qquad \mbox{and} \qquad 
\twocell{(1 *0 d) *1 (1 *0 a) *1 (1*0 d) *1 ln *1 (d*0 1)} =  \twocell{ (1*0 d) *1 ln *1 (d*0 1) *1 (a*0 1)*1 (d *0 1)} \quad \mbox{ for all } \alpha \in \s_2 $$
in the $2$-category $\s^*$.

\end{defi}

\begin{teor}[Convergence of $\mathfrak S$]\label{permconf}
The polygraph $\mathfrak S$ is convergent.
\begin{proof}
As in \cite{LafBool}, in order to prove termination we interprete every diagram $\phi: n\fr m \in  \mathfrak S^*$ with a monotone function $[\phi]:\N^n\fr \N^m$. These have a well founded order induced by product order on $\N^p$:
$$f,g: \N^{*p}\fr \N^{*p} \mbox{ then } f\geq g \mbox { iff } f(\bar x)\geq g(\bar x) \mbox{ for all } \bar x \in \N^{*p}.$$
We interprete the gate $\swap $ by the function $[\;\twocell{s}\;] (x,y)\fr (y, x+y)$.
This allow as to associate to any $3$-cell $\xymatrix@C-1pc{ \phi \ar@3[r] & \psi}$ two monotone maps  $[\phi]$ and $[\psi]$ such that $[\phi]> [\psi]$:
$$\begin{gathered}
\Big[\; \twocell{s *1 s }\;\Big] (x,y)= (2x+y,x+y)> (x,y)=\Big[\;\twocell{2}\;\Big](x,y), \\
\Bigg[\;{\twocell{(s *0 1)*1 (1 *0 s)*1 (s *0 1)}}\;\Bigg](x,y,z)=(2x+y+z, x+y,x)> (x+y+z,x+y,x)=\Bigg[\;{\twocell{ (1 *0 s)*1 (s *0 1)*1 (1 *0 s) }}\;\Bigg](x,y,z)
\end{gathered}$$
By the compatibility of the order with sequential and parallel composition, this suffice to prove that, for any couple of  diagrams, $[\phi] > [\psi]$ holds if $\phi\fr^* \psi$. Since there exists no infinite decreasing suite of monotone maps on positive integers, infinite reduction paths can not exist.

In order to prove convergence, it suffices to check the confluence of the following critical peaks, that are minimal critical branchings (see \cite{AccCohe}, App.A for details):
$$
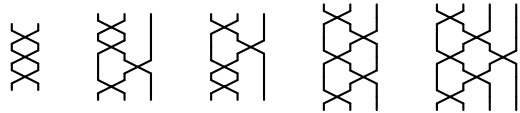
\twocell{s *1 s *1 s}								\quad \quad
\twocell{(s *0 1) *1 (s *0 1)*1 (1 *0 s)*1 (s *0 1)} 				\quad \quad
\twocell{(s *0 1)*1 (1 *0 s)*1 (s *0 1) *1 (s *0 1)} 				\quad \quad
\twocell{(s *0 1) *1 (1*0 s) *1 (s *0 1)*1 (1 *0 s)*1 (s *0 1)} 		\quad \quad
\twocell{(s *0 2) *1 (1*0 s *0 1) *1 (s *0 s)*1 (1 *0 s *0 1) *1 (s *0 2)}	
$$  
\end{proof}
\end{teor}

Each diagram in $\mathfrak S$ can be interpreted as a permutation in the \emph{group of permutations over $n$ elements} $S_n$ with product $\circ$ defined as their function composition. On the other hand, each $\sigma\in S_n$ corresponds to some diagrams in $\mathfrak S$. In particular, we interpret the diagram $\id_{k-1} * \swap * \id_{n-(k+1)}: n \fr n$ as the transposition $(k,k+1)\in S_n$.

\begin{prop}\label{corrPer}
For any permutation $\sigma\in S_n$ there is a unique diagram in normal form $\hat \phi_\sigma: n\frr n \in \mathfrak S$ corresponding to $\sigma$. We call it the \emph{canonical diagram of $\sigma$}.
\begin{proof}
We define $\mathfrak S_1=\{\twocell{1}\}$ and $\mathfrak S_{n+1}$ the set of diagrams in $ \mathfrak S$ of the form:
$$\twocell{(1 *0 d) *1 (1 *0 sigma1) *1 (1 *0 d) *1 (g35) *1 (d *0 1 *0 d)}=\hat\phi_\sigma: n+1\frr n+1 $$
with $\twocell{ d *1 sigma1 *1 d} \in \mathfrak S_{n}$ and $\twocell{(1 *0 d)*1 g35 *1 (d *0 1 *0 d)} = \twocell{(1 *0 d *0 2)*1 (ln *0 d) *1( d *0 3)}= Lad^l_k * id_{(n+1-k)}$. 
We have $|\mathfrak S_n|=n!$ since $|\mathfrak S_1|=1$ and  $|\mathfrak S_{n+1}|=(n+1)|\mathfrak S_{n}| $ on account of $n+1=|\{Lad^l_k\}_{1\leq k\leq n+1}|=|\{Lad^l_k* \id_{(n+1-k)}\}_{1\leq k\leq n+1}|$.

To exhibit a one-to-one correspondence between $S_{n+1}$ and $\mathfrak S_{n+1}$, for any $\sigma \in S_{n+1}$ we define $Er(\sigma)\in S_n$ the permutation
$$Er(\sigma)=\begin{cases}i \fr \sigma(i+1) & {if }\quad \sigma (i+1)<\sigma (1) \\ i \fr \sigma(i+1)+1 & {if }\quad \sigma (1)<\sigma (i+1)\end{cases}.$$
and $\hat \phi _\sigma=( Lad^l_k* \id_{(n+1-\sigma(1))}) \circ (\id_1 * \hat \phi _{Er(\sigma)})$.

Any element in $\mathfrak S_n$ contains no subdiagram of the form $\twocell{s *1 s}$ nor $\twocell{(s *0 1) *1 (1 *0 s) *1 (s *0 1)}$ meaning that  it is irreducible and so, by the confluence of $\mathfrak S$, in normal form.
\end{proof}
\end{prop}

\nota  We note $ \twocell{ (1*0 d) *1 ln *1 (d*0 1) }: n\frr n $ and $ \twocell{(d*0 1) *1 rn *1 (1*0 d) }:n\frr n$ the diagrams corresponding respectively to the permutations $(1,n,n-1, \dots , 2)$ and $(n,1,2, \dots, n-1)$ in $S_n$.

\begin{defi}[Twisting polygraph]
A \emph{twisting polygraph} is  a $3$-polygraph $\s$ with one $0$-cell equipped with a set $T_\s\subseteq \s_1 $ called \emph{twisting family} such that  for each $A,B \in T_\s$ there is a \emph{twisting operator} $\swapl A {B}: A *B \frr B*A \in \s_2$  and $\s_3$  includes the following families $T_\R$ of \emph{twisting relations}:
\begin{itemize}
\item For all $A,B,C \in T_\s$:
\vspace{-.4 cm}
\begin{equation}\label{twist1}\xymatrix{ {\twocell{(topA *0 topB )*1(s *1 s) }} \ar@3[r] & {\twocell{(midA *0 midB)*1 2}}} \qquad \mbox{ and }\qquad 
\xymatrix{ \twocell{(topA *0 topB *0 topC)*1(s *0 1)*1 (1 *0 s)*1 (s *0 1)*1(pitC *0 pitB *0 pitA)} \ar@3[r] &{\twocell{ (topA *0 topB *0 topC)*1(1 *0 s)*1 (s *0 1)*1 (1 *0 s) *1(pitC *0 pitB *0 pitA) }}} 
\quad ; \end{equation}
\vspace{-.6 cm}
\item For all $\alpha: \Gamma \fr \Gamma' \in \s_2$ with $\Gamma, \Gamma'\in T_\s^*$, $A\in T_\s$, at least one of the two possible orientation of the following rewriting rules is in $\s_3$:
\vspace{-.4 cm}
\begin{equation}\label{twist2}
\xymatrix{ \twocell{(topGam *0 topA )*1 (d *0 1) *1 (a *0 1) *1 (d*0 1) *1 rn *1 (1*0 d) *1 (pitA *0 pitGam1)}  \ar@3@/^1pc/[r] &  
\twocell{ (topGam *0 topA)*1(d*0 1) *1 rn *1 (1*0 d) *1 (1*0 a)*1 (1 *0 d)*1 (pitA *0 pitGam1)} \ar@3@/^1pc/[l]}
\qquad \mbox{and} \qquad
\xymatrix{ \twocell{ (topA *0 topGam)*1 (1 *0 d) *1 (1 *0 a) *1 (1*0 d) *1 ln *1 (d*0 1) *1 (pitGam1 *0 pitA)} \ar@3@/^1pc/[r] & \twocell{(topA *0 topGam)*1 (1*0 d) *1 ln *1 (d*0 1) *1 (a*0 1)*1 (d *0 1)*1 (pitGam1 *0 pitA)}\ar@3@/^1pc/[l] }
\quad.
\end{equation}\vspace{-.6 cm}
\end{itemize}
Moreover, if $\phi, \psi$ are \emph{twisting diagrams} (i.e. diagrams made only of twisting operators) $\xymatrix@C=1em{\phi \ar@3[r]^{*}_{~\R_\s} & \psi}$ iff $\xymatrix@C=1em{\phi \ar@3[r]^{*}_{~\R_T} & \psi}$ where $\R_T$ is the set given by rewriting rules of (\ref{twist1}).
A \emph{total-twisting polygraphy} is a twisting polygraph with $T_\s=\s_1$.
\end{defi}

The idea  behind twisting polygraphs is to present diagram rewriting systems where, in equivalence classes modulo rewriting, the crossings of strings labeled by the twisting family are not taken into account. In fact, the family of relations (\ref{twist1}) says that these crossings are involutive and satisfy Yang-Baxter equation \cite{YBE} for braidings, while relations in (\ref{twist2}) allow gates to ``cross'' a string in case of fitting  labels. 

We interpret a twisting diagram $\phi_\sigma :\Gamma \frr \sigma(\Gamma)$ as the permutations  in $S_{|\Gamma|}$ acting over the order of occurrence of $1$-cells in the word $\Gamma \in T_\s^*$. For this reason, as in $\mathfrak S$, we define left ladders, right ladders and the standard diagrams $\hat \phi^\Gamma_\sigma : \Gamma \fr \sigma(\Gamma)$ (or simply $\hat \phi _\sigma$)  with source and target in $T_\s^*$. In conformity with the twisting polygraph restrictions over $\s_3$, we can prove the uniqueness of $\hat \phi_\sigma$ as in Proposition \ref{corrPer}.

\section{Multiplicative Linear Logic sequent calculus}

In this paper we focus on the multiplicative fragment of linear logic sequent calculus with or without units.
We here we recall the usual inference rules:

{\small\begin{tabular}
{p{2,5cm} |p{4cm}|p{4cm}}
&  Identity or Axiom  &  Cut \\
Structural &
{\begin{prooftree}
\AxiomC{}
\RightLabel{$Ax$}
\UnaryInfC{$ \vdash A, A^\bot$}
\end{prooftree}}
&
{\begin{prooftree}
\AxiomC{$\vdash \s , A$}
\AxiomC{$\vdash \Gamma, A^\bot$}
\RightLabel{$Cut$}
\BinaryInfC{$ \vdash \s , \Gamma$}
\end{prooftree}}\\
\hline
&  Tensor & Par\\
 Multiplicative &
\begin{prooftree}
\AxiomC{$\vdash \s , A$}
\AxiomC{$\vdash B , \Gamma $}
\RightLabel{$\otimes $}
\BinaryInfC{$ \vdash \s,  (A\otimes B), \Gamma $}
\end{prooftree}
&
\begin{prooftree}
\AxiomC{$\vdash \s, A,B $}
\RightLabel{$\parr $}
\UnaryInfC{$ \vdash \s , A\parr B$}
\end{prooftree}\\
\hline
& Bottom & 1\\
Units &
 \begin{prooftree}
\AxiomC{$\vdash \s $}
\RightLabel{$\bot $}
\UnaryInfC{$ \vdash \s ,\bot$}
\end{prooftree}
&
\begin{prooftree}
\AxiomC{$  $}
\RightLabel{$1 $}
\UnaryInfC{$ \vdash \mathit {1}$}
\end{prooftree}
\end{tabular}}

We also consider the usually omitted exchange rule: 
\begin{prooftree}
\AxiomC{$\vdash A_1, \dots , A_k $}
\RightLabel{$ \sigma \in S_k $}
\UnaryInfC{$\vdash A_{\sigma(1)}, \dots , A_{\sigma(k)}$}
\end{prooftree}

We finally recall that the \emph{multiplicative linear logic fragment with units} ($\MLLc$) is given by the aforementioned inference rules while the \emph{multiplicative fragment} ($\MLL$) is the one given by the inference rules $Ax, Cut, \otimes , \parr$ (and exchange) only.

\begin{oss}[On Negation]
We assume negation is involutive, i.e. $A^{\bot \bot}=A$ and the De-Morgan laws apply with respect to $\parr$ and $\otimes$, i.e. $(A \heartsuit B)^\bot= B^\bot \heartsuit^\bot A^\bot$ for any formulas $A,B$ where $\heartsuit=\parr$ and $\heartsuit^\bot=\otimes$ or vice versa $\heartsuit= \otimes $ and $\heartsuit^\bot= \parr$. Moreover $1^\bot=\bot$.
\end{oss}

\begin{oss}[On Rules]\label{arity}
In this work we interpret inference rules as operations with specific arities over the set of sequents: $Ax$ and $1$ are $0$-ary, $\parr$ and $\bot$ are unary and $\otimes$ and $Cut$ are binary.
\end{oss}

\nota We indicate with $\fmll$ and $\fmllc$ the set of formulas respectively in $\MLL$ and $\MLLc$.

\vspace{-.3cm}

\section{String diagram syntax for proof net}
\vspace{-.3cm}
In this section we give two particular $3$-polygraphs for $\MLL$ and $\MLLc$ respectively,  i.e. string diagrams representing linear logic derivations that we call \emph{proof diagrams}.
To these latter, we then add two non-twisting colors and we replace certain $2$-cells in order to define what we call  \emph{control polygraphs}. In these polygraphs we are able to characterize diagrams corresponding to correct proof structures by just checking their inputs and outputs patterns.

\nota In order to unify sequent and $1$-cell composition notations, we replace the $*$ symbol of parallel composition with a comma.

\vspace{-.2cm}
\subsection{Proof diagrams for $\MLL$}
 

\begin{defi}
The $3$-polygraph $\s_{MLL}$ is the \emph{polygraph of multiplicative linear logic with cut-elimination}. It is given by the following sets of cells:
\begin{multicols}{2}
\begin{itemize}
\item $\s^M_0=\{\pro\}$;
\item $\s^M_1=\fmll $;
\end{itemize}
\end{multicols}
\begin{itemize}
\item $\s^M_2=\begin{Bmatrix}
\otimes_{A,B}: &  A , B &\frr & A\otimes B& = & \twocell{(topA *0 topB) *1 ten *1 (pitAtenB)}  \\
\parr_{A,B}: &  A , B &\frr & A\parr B  & = &      \twocell{(topA *0 topB) *1 par *1 (pitAparB)} \\
Ax_A: & \pro &\frr & A , A^\bot  & = & {\twocell{axA *1 (pitA *0 pitAb)}}\\
Cut_A: &  A , A^\bot &\frr & \pro  & = & {\twocell{(topA *0 topAb) *1 cutA}}\\
\swapl {A}{B}: &  A , B &\frr & B , A & = & \twocell{(topA *0 topB)*1 (s)*1(pitB *0 pitA)}
\end{Bmatrix}_{A,B\in \fmll}$

If there is no ambiguity we note \twocell{ax} and \twocell{cut} instead of \twocell{axA} and \twocell{cutA}.

\item $\s^M_3=\s^M_{Twist} \cup \s^M_{Cut}$ where:

\begin{itemize} 
\item $\s^M_{Twist}$ is given by the following twisting relations:
\vspace{-.5 cm}
\end{itemize}
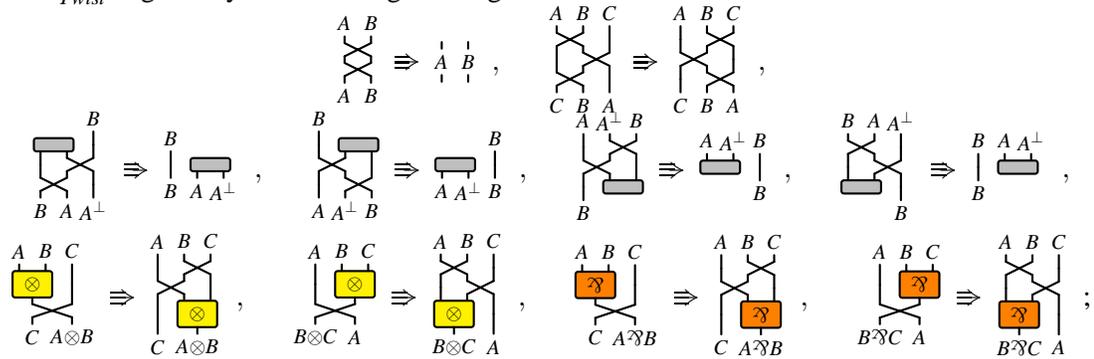
$$
\xymatrix@C=.3cm{ {\twocell{(topA *0 topB )*1(s *1 s)  *1 (pitA *0 pitB)}} \ar@3[r] & {\twocell{(midA *0 midB)}}} ,
\quad 
\xymatrix@C=.3cm{ \twocell{(topA *0 topB *0 topC)*1(s *0 1)*1 (1 *0 s)*1 (s *0 1)*1(pitC *0 pitB *0 pitA)} \ar@3[r] &{\twocell{ (topA *0 topB *0 topC)*1(1 *0 s)*1 (s *0 1)*1 (1 *0 s) *1(pitC *0 pitB *0 pitA) }}},
$$
\vspace{-.8 cm}
$$
\xymatrix@C=.3cm{ \twocell{topB *1 (ax *0 1) *1 (1 *0 s)*1 (s *0 1) *1 (pitB *0 pitA *0 pitAb)} \ar@3[r] & \twocell{topB *1 ( 1 *0 ax)*1 (pitB *0 pitA *0 pitAb) }},
\quad
\xymatrix@C=.3cm{  \twocell{topB *1 (1*0 ax )*1 (s*0 1) *1 (1*0 s) *1 (pitA *0 pitAb *0 pitB)} \ar@3[r] & \twocell{ topB *1 (ax *0 1 ) *1 (pitA *0 pitAb *0 pitB)}},
\quad 
\xymatrix@C=.3cm{\twocell{(topA *0 topAb *0 topB ) *1 ( 1 *0 s)*1 (s *0 1)*1 (1*0 cut)*1pitB} \ar@3[r] &  \twocell{(topA *0 topAb *0 topB ) *1( cut *0 1) *1 pitB }},
\quad
\xymatrix@C=.3cm{ \twocell{(topB *0 topA *0 topAb) *1 (s*0 1) *1 (1*0 s) *1 (cut *0 1)*1 pitB} \ar@3[r] &  \twocell{ (topB *0 topA *0 topAb) *1 (1 *0 cut) *1 pitB }},
$$
\vspace{-.6 cm}
$$ 
\xymatrix@C=.3cm{\twocell{(topA *0 topB *0 topC) *1 (ten *0 1) *1  s *1(pitC *0 pitAtenB) } \ar@3[r] & \twocell{(topA *0 topB *0 topC) *1 (1 *0 s)*1 (s *0 1)*1 (1 *0 ten)*1(pitC *0 pitAtenB)}},
\quad
\xymatrix@C=.3cm{ \twocell{(topA *0 topB *0 topC) *1 (1 *0 ten)*1 (s) *1 (pitBtenC *0 pitA)} \ar@3[r] & \twocell{(topA *0 topB *0 topC) *1 (s *0 1)*1 (1 *0 s)*1 (ten *0 1)*1 (pitBtenC *0 pitA) }},
\quad
\xymatrix@C=.3cm{\twocell{(topA *0 topB *0 topC) *1(par *0 1)*1 (s)*1 ( pitC *0 pitAparB)} \ar@3[r] & \twocell{(topA *0 topB *0 topC) *1 (1 *0 s)*1 (s *0 1)*1 (1 *0 par)*1 ( pitC *0 pitAparB) }},
\quad
\xymatrix@C=.3cm{\twocell{(topA *0 topB *0 topC) *1( 1 *0 par)*1 (s)*1(pitBparC *0 pitA)} \ar@3[r] & \twocell{(topA *0 topB *0 topC) *1 (s *0 1)*1 (1 *0 s)*1 (par *0 1)*1 (pitBparC *0 pitA) }};
$$\vspace{-.7 cm}

together with two rules representing the involution $A^{\bot\bot}=A$: $
\xymatrix{ {\underset{~A\; ~A^\bot}{\twocell{ axA *1 s }}} \ar@3[r] & {\underset{A^\bot\;  A}{\twocell{ axAb }}}},
\;
\xymatrix{ {\overset{~A\; ~A^\bot}{\twocell{ s *1 cutA }}} \ar@3[r] & {\overset{A^\bot\;  A}{\twocell{ cutAb }}}};
$
\vspace{-.3 cm}
\begin{itemize} 
\item $\s^M_{Cut}$ is the set of rules for the cut elimination:
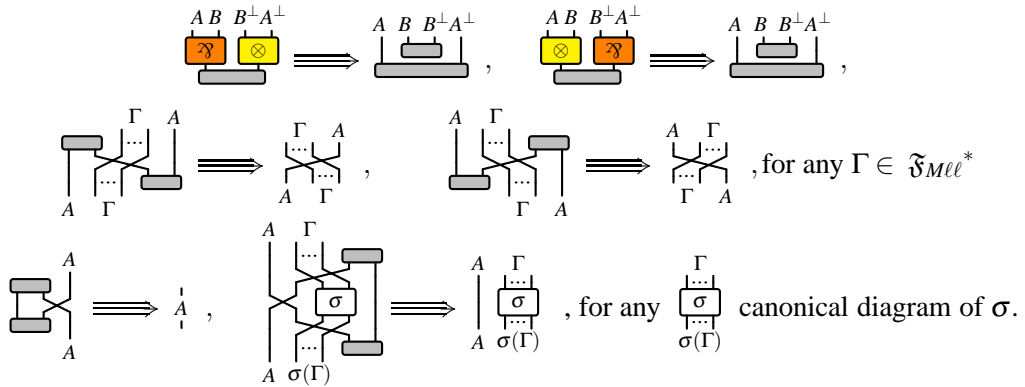
\vspace{-.3 cm}
$$
\xymatrix{ \overset{~~A\; B\; ~B^\bot A^\bot}{\twocell{(par *0 ten)*1 cut }} \ar@3[r] & \overset{~A\; ~B\;~ B^\bot A^\bot}{\twocell{ (1 *0 cut *0 1) *1 ( cut)}}},
\quad
\xymatrix{ \overset{~~A\; B\; ~B^\bot A^\bot}{\twocell{(ten *0 par)*1 cut }} \ar@3[r] & \overset{~A\; ~B\;~ B^\bot A^\bot}{\twocell{ (1 *0 cut *0 1) *1 ( cut)}}},\\
$$
\vspace{-.6 cm}
$$
\xymatrix{ \twocell{(topGam *0 topA) *1 (ax *0 d *0 1) *1 (1 *0 ln *01) *1 (1*0 d*0 cut)*1 (pitA *0 pitGam) } \ar@3[r] & \twocell{(topGam *0 topA) *1 (d*0 1) *1 rn *1 (1*0 d)*1 (pitA *0 pitGam)}},
\qquad 
\xymatrix{  \twocell{(topA *0 topGam)*1 (1 *0 d *0 ax) *1 (1 *0 rn *01) *1 (cut*0 d*0 1) *1 (pitGam *0 pitA) } \ar@3[r] & \twocell{(topA *0 topGam) *1 (1*0 d) *1 ln*1 (d *0 1)*1 (pitGam *0 pitA)}}, \mbox{for any $\Gamma\in\ \fmll^*$}
$$
\vspace{-.7 cm}
$$
\xymatrix{ \twocell{topA *1 (ax *0 1) *1 (1*0 s) *1 (cut *0 1) *1 pitA } \ar@3[r] & \twocell{midA}},
\quad 
\xymatrix{ \twocell{(topA  *0 topGam)  *1 (1 *0 d *0 ax) *1  (1 *0 rn *0 1)*1 (s *0 sigma *0 1)*1 (1*0 ln*0 1) *1 (1*0 d*0 cut) *1 (pitA *0 pitsiG)} \ar@3[r] & \twocell{(topA  *0 topGam)  *1 (1 *0 d)*1 (1 *0 sigma) *1 (1 *0 d)*1 (pitA *0 pitsiG) }} \; \mbox{, for any $\twocell{topGam *1 d *1 sigma *1 d *1 pitsiG}$ canonical diagram of $\sigma$.}
$$

\end{itemize}
\end{itemize}
\vspace{-.4 cm}
\end{defi}

\begin{teor}[Interpretation of proofs in $\s_{MLL}$]\label{intmll}
For any derivation $d(\Gamma)$ of $\vdash \Gamma$ in $\MLL$ there is a proof diagram $\phi_{d(\Gamma)}: \pro \frr \Gamma\in \s_{MLL}$.
\begin{proof}
Let $d(\Gamma)$ be a derivation in $\MLL$ of $\vdash \Gamma$. First we observe that, if there is a diagram $\phi: \Delta \frr \Gamma$ so there is a diagram $\phi^\sigma= \hat \phi_\sigma \circ \phi: \Delta \frr \sigma(\Gamma)$ for all permutation $\sigma\in S_{|\Gamma|}$. By this fact we can proceed by induction on the number of inference rules appearing in $d(\Gamma)$:
\begin{itemize}
\item If just one inference rule occurs in $d(\Gamma)$, it must be an $Ax$ rule, $\Gamma=A,A^\bot$ and $\phi_{d(\Gamma)}=Ax_A:\pro \frr A, A^\bot$;

\item If $n+1$ inference rules occur in $d(\Gamma) $, then we consider the last one and we distinguish two cases in base of its arity (see Rem. \ref{arity}):
\begin{itemize}

\item If it is unary and $ \Gamma= \Gamma', A\parr B$, then, by inductive hypothesis, there is a diagram $\phi_{d(\Gamma',A,B)}: \pro \fr \Gamma',A,B$ of the derivation $d( \Gamma',A,B)$ with $n$ inference rules. Therefore
$$\phi_{d(\Gamma)}=(\id_{\Gamma'},\parr_{A,B}) \circ \phi_{d(\Gamma',A,B)}: \pro \frr \Gamma ;$$

\item If it is binary and $ \Gamma=\Delta, A\otimes B, \Delta'$, then, by inductive hypothesis, there are two diagrams $\phi_{d(\Delta, A)}:\pro \frr \Delta, A$ and $\phi_{d(B, \Delta')}:\pro \frr B ,\Delta'$ relative to the two derivations $d(\Delta, A)$ and $d(B, \Delta')$ with at most $n$ inference rules. Therefore
$$\phi_{d(\Gamma)}=(\id_{\Delta},\otimes_{A,B}, \id_{\Delta'}) \circ (\phi_{d(\Delta, A)} , \phi_{d(B, \Delta')}):  \pro \frr \Gamma ;$$

\item Similarly, if it is binary and $\Gamma= \Delta, Cut(A, A^\bot), \Delta'$, then
$$\phi_{d(\Gamma)}=(\id_{\Delta},cut_{A}, \id_{\Delta'}) \circ (\phi_{d(\Delta, A)} , \phi_{d(A^\bot, \Delta')}):  \pro \frr \Gamma . $$
\end{itemize}
\end{itemize}
\vspace{-.5cm}
\end{proof}
\end{teor}


\subsection{Proof diagram with control for $\MLL$}
In order to have a correctness criterion for $\MLL$ proof diagrams, we enrich the set of string labels with two new non-twisting colors $L$ (left) and $R$ (right) and re-define some $2$-cells. 

The idea is to use these latter to introduce a notion of well-paranthesization in a setting where a proof derivation can be seen as a sequence of operations over lists of sequents: unary derivation rules act on single sequents (as in the case of $\parr$), binary ones act on two sequent (as in the case of $\otimes$ and $Cut$) and the 0-ary one, that is $Ax$, generates a new sequent.


\begin{defi}
The \emph{control polygraph of multiplicative linear logic} $\Mllt$ is given by the following sets of cells:
\begin{multicols}{2}
\begin{itemize}
\item $\Mllt_0=\{\pro\}$;
\item $\Mllt_1= \fmll \cup \{L=\twocell{L},R=\twocell{R}\}$;
\end{itemize}
\end{multicols}
\begin{itemize}
\item $\Mllt_2=\begin{Bmatrix}
\otimes_{A,B}: &  A ,R,L, B &\frr & A\otimes B &=& \twocell{(topA *0 R *0 L *0 topB)*1 tenc *1 pitAtenB}\\
\parr_{A,B}: &  A , B &\frr & A\parr B  &=& \twocell{(topA *0 topB )*1 par *1 pitAparB}\\
Ax_A: & \pro &\frr &L, A , A^\bot,R &=& \twocell{axcA*1 (L*0 pitA *0 pitAb*0 R)} \\
Cut_A: &  A ,R,L, A^\bot &\frr & \pro  &=& \twocell{(topA *0 R *0 L *0 topAb)*1 cutc}\\ 
\swapl {A}{B}: &  A ,B &\frr & B , A &=& \twocell{(topA *0 topB) *1 s *1 (pitB *0 pitA)}\\
\end{Bmatrix}_{A,B\in \fmll}$

\item $\Mllt_3= \Mllt_{Twist}$  is given by the  following twisting relations:
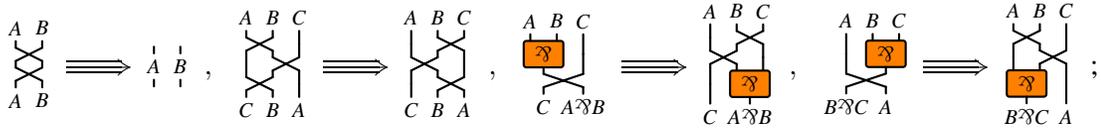
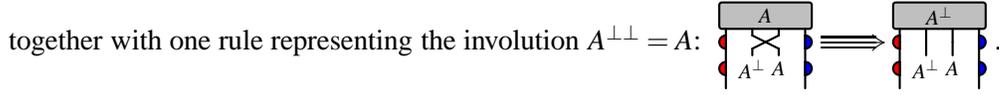
\vspace{-.5 cm}
$$
\xymatrix{ {\twocell{(topA *0 topB )*1(s *1 s) *1 (pitA *0 pitB) }} \ar@3[r] & {\twocell{(midA *0 midB)}}}, 
\; 
\xymatrix{ \twocell{(topA *0 topB *0 topC)*1(s *0 1)*1 (1 *0 s)*1 (s *0 1)*1(pitC *0 pitB *0 pitA)} \ar@3[r] &{\twocell{ (topA *0 topB *0 topC)*1(1 *0 s)*1 (s *0 1)*1 (1 *0 s) *1(pitC *0 pitB *0 pitA) }}} ,
\; 
\xymatrix{\twocell{(topA *0 topB *0 topC) *1(par *0 1)*1 (s)*1 ( pitC *0 pitAparB)} \ar@3[r] & \twocell{(topA *0 topB *0 topC) *1 (1 *0 s)*1 (s *0 1)*1 (1 *0 par)*1 ( pitC *0 pitAparB) }},
\;
\xymatrix{\twocell{(topA *0 topB *0 topC) *1( 1 *0 par)*1 (s)*1(pitBparC *0 pitA)} \ar@3[r] & \twocell{(topA *0 topB *0 topC) *1 (s *0 1)*1 (1 *0 s)*1 (par *0 1)*1 (pitBparC *0 pitA) }};
$$\vspace{-.5 cm}

together with one rule representing the involution $A^{\bot\bot}=A$: $\xymatrix{ \twocell{ axcA *1 (L *0  s *0 R) *1 (L *0 pitAb *0 pitA *0 R) } \ar@3[r] & \twocell{ axcAb *1 (L *0 2 *0 R) *1 (L *0 pitAb *0 pitA *0 R) }}$.

\end{itemize}
\end{defi}


\begin{oss}
The polygraph $\Mllt$ is twisting with twisting family $\fmll$.
\end{oss}

\begin{teor}[Proof diagrams correspondence in $\Mllt$]\label{corrMLL}
$$\vdash_{\MLL}\Gamma \Leftrightarrow \exists \phi \in \Mllt \mbox{ such that } \phi: \pro \frr L,\Gamma, R.$$
\begin{proof}
To prove the left-to-right implication $\Rightarrow$, as in Teor. \ref{intmll}, we remark that, if there is a diagram $\phi: \pro \frr L, \Gamma, R$ with $\Gamma$ sequent in $\MLL$, so there is a diagram 
$$\phi^\sigma= (\id_L ,\hat \phi_\sigma, \id_R) \circ \phi: \pro \frr L,\sigma(\Gamma), R $$
 for any permutation $\sigma\in S_{|\Gamma|}$. Then we proceed by induction on the number of inference rules in a derivation $d(\Gamma)$ in $MLL$:
\begin{itemize}
\item If just one inference rule occurs $d(\Gamma)$, then it is an $Ax$ and $\Gamma=A,A^\bot$ and $\phi_{d(\Gamma)}=Ax_A: \pro \frr L,A, A^\bot , R$;

\item If $n+1$ inference rules appear, then we consider the last one and we distinguish two cases in base of its arity:
\begin{itemize}

\item If it is an unary $\parr$ and $ \Gamma= \Gamma', A\parr B$, then, by inductive hypothesis, there is a diagram $\phi_{d(\Gamma',A,B)}: \pro \frr L, \Gamma',A,B, R$ of the derivation $d( \Gamma',A,B)$ and 
$$\phi_{d(\Gamma)}=(\id_{L,\Gamma'},\parr_{A,B},\id_R) \circ \phi_{d(\Gamma',A,B)}: \pro \frr L, \Gamma, R ;$$

\item If it is a binary $\otimes$ and $ \Gamma=\Delta, A\otimes B, \Delta'$, then, by inductive hypothesis, there are two diagrams $\phi_{d(\Delta, A)}:\pro \frr L, \Delta, A, R$ and $\phi_{d(B, \Delta')}:\pro \frr L, B ,\Delta', R$ relative to the two derivations $d(\Delta, A)$ and $d(B, \Delta')$ with at most $n$ inference rules. Therefore
$$\phi_{d(\Gamma)}=(\id_{L,\Delta},\otimes_{A,B}, \id_{\Delta',R}) \circ (\phi_{d(\Delta, A)} , \phi_{d(B, \Delta')}): \pro \frr L, \Gamma, R $$

\item Similarly, if it is a binary $Cut$ and $\Gamma= \Delta, Cut(A, A^\bot), \Delta'$, then
$$\phi_{d(\Gamma)}=(\id_{L,\Delta},Cut_{A^\bot}, \id_{\Delta',R}) \circ (\phi_{d(\Delta, A)} , \phi_{d(A^\bot, \Delta')}): \pro \frr L, \Gamma, R. $$

\end{itemize}
\end{itemize}

In order to prove sequentialization, i.e. the right-to-left implication $\Leftarrow$, we proceed by induction on the number $|\phi|_\sig$ of gates in $\phi$:
\begin{itemize}
\item If $|\phi|_{ \Mllt}=0$ so $\phi: \id_\Gamma: \Gamma \frr \Gamma$. By hypothesis $\phi$ has no input (i.e. $s_2(\phi)=\pro$) so it is the identity  diagram over the empty string, this is the empty diagram $\id_0: \pro\frr \pro$ which it is not sequentializable since $t_2(\phi)=\pro \neq L,R$;
\item If  $|\phi|_{ \Mllt}=1$ than $\phi$ is an elementary diagram. The elementary diagrams with source $\pro$ and target $L,\Gamma, R$ with $\Gamma \in \fmll^*$ are atomic made of a unique $2$-cell of type $Ax_A: \pro \frr L, A, A^\bot, R$ for some  $A\in \fmll$. The associated sequent $\vdash A, A^\bot$ is derivable in $\MLL$;

\item Otherwise there is  $2$-cell of type  $\alpha: \Gamma' \frr \alpha ( \Gamma' ) \in \Mllt_2$ and  $\Gamma= \Delta, \alpha (\Gamma' ), \Delta'$. In this case $\phi= (\id_{L, \Delta} , \alpha , \id_{\Delta, R}) \circ \phi' $ where $\phi': \pro \frr L,\Delta, \Gamma', \Delta',R$. We have the following cases:

\begin{itemize}
\item If $\alpha= \swapl A B $, $\Gamma' = A,B$ and $\alpha(\Gamma')= B,A$. The diagram $\phi'$ is sequentializable by inductive hypothesis since $|\phi|_{\Mlltc}= |\phi' |_{\Mllt} +1$;
\item Similarly if  $\alpha= \parr_{A,B}$, $\Gamma'= A,B$ and $\alpha(\Gamma')= A\parr B$;
\item If $\alpha= \otimes_{A,B}$ so $\Gamma'= A, R, L, B$,  $\alpha(\Gamma')= A\otimes B$ and 
$$\phi': \pro \frr L,\Delta, A, R, L, B , \Delta',R. $$
This diagram is a parallel composition $\phi=\phi'_l ,\phi'_r$ with 
$$\phi'_l : \pro \frr L,\Delta, A, R\quad \mbox{ and } \quad  \phi'_r : \pro \frr L,B, \Delta',  R$$
of two diagrams which satisfy
 inductive hypothesis since $|\phi|_{\Mllt}=|\phi'_l|_{\Mllt} +|\phi'_r|_{ \Mllt}+1$;
\item Similarly if $\alpha= Cut_{A}$ with $B=A^\bot$ we have $\Gamma'=A,R, L,  A^\bot$ and $\alpha(\Gamma')=\emptyset$. 
\end{itemize}
\end{itemize}
\vspace{-.7cm}
\end{proof}
\end{teor}


\subsection{Proof diagrams for $\MLLc$}
In this section we extend the signatures of the two previous polygraphs  in order to accommodate multiplicative units in our syntax of proof diagrams and we enunciate some relation between this syntax and the multiplicative proof structure's one.


\begin{defi}
The \emph{polygraph of multiplicative linear logic with constants and cut-elimination} $ \s_{MLLc}$ is given by the following sets of cells:
\begin{multicols}{2}
\begin{itemize}
\item $\s^u_0=\{\pro \}$;
\item $\s^u_1=\fmllc$;
\end{itemize}
\end{multicols}
\begin{itemize}
\item $\s^u_2=\begin{Bmatrix}
1: & \pro &\frr & 1 & =& \twocell{v *1 pit1}\\
\bot: & \pro &\frr &\bot  &=& \twocell{bot *1 pitbot}\\
\end{Bmatrix}\cup \s^M_2$
\item $\s_3^u=\s^u_{Twist} \cup \s^{u}_{Cut}$ where:
\begin{itemize} 
\item $\s^u_{Twist}$ is $\s^M_{Twist}$ along with the following twisting relations:\vspace{-.3 cm}
$$
\xymatrix{ \twocell{topA *1 (bot *0 1) *1  s *1( pitA *0 pitbot) } \ar@3[r] & \twocell{ topA *1 (1 *0 bot)  *1( pitA *0 pitbot) }},
\quad
\xymatrix{ \twocell{topA *1 (1 *0 bot)*1 (s) *1 (pitbot *0 pitA)} \ar@3[r] & \twocell{ topA *1 (bot *0 1) *1 (pitbot *0 pitA) }},
\quad
\xymatrix{ \twocell{topA *1 (v *0 1) *1  s *1( pitA *0 pit1) } \ar@3[r] & \twocell{topA *1  (1 *0 v) *1( pitA *0 pit1)}},
\quad
\xymatrix{ \twocell{topA *1 (1 *0 v)*1 (s) *1 (pit1 *0 pitA)} \ar@3[r] &\twocell{ topA *1 (v *0 1)*1 (pit1 *0 pitA) }};
$$\vspace{-.3 cm}

\item  $\s^{u}_{Cut}$ is $\s^M_{Cut}$ along with the following rules for cut elimination:
$
\xymatrix{ {\twocell{(bot *0 v)*1 cut }} \ar@3[r] & \emptyset },
\;
\xymatrix{ {\twocell{(v *0 bot)*1 cut }} \ar@3[r] & \emptyset }
$.
\end{itemize}

\end{itemize}
\end{defi}

\begin{oss}
The polygraph $\s_{MLLu}$ is total-twisting.
\end{oss}

\begin{teor}[Interpretation of proofs in $\s_{MLLu}$]\label{intmllc}
For any derivation $d(\Gamma)$ of $\vdash \Gamma$ in $\MLLc$ there is a proof diagram $\phi_{d(\Gamma)}: \pro \frr \Gamma\in \s_{MLLu}$.
\begin{proof}
The proof is much like the one we provided for Theorem \ref{intmll} . In order to accommodate units, we just need to slightly revisit our inductive reasoning by considering the following two additional cases (i.e. the remaining cases stay the same):
\begin{itemize}
\item If just one inference rule occurs in $d(\Gamma)$, then it may be a $1$ rule (in addition to $Ax$). It follows that $\Gamma=1$ and $\phi_\Gamma= 1: \pro \frr 1$;

\item If the last of the $n+1$ inference rules appearing in $d(\Gamma) $ is an unary $\bot$ and $\Gamma= \Gamma', \bot$, then, by inductive  hypothesis, there is a  diagram $\phi_{\Gamma'}: \pro \frr \Gamma'$ and $\phi_\Gamma= \phi_{\Gamma'}, \bot$.
\end{itemize}\vspace{-.6cm}
\end{proof}
\end{teor}





In the extended version of this paper\footnote{https://arxiv.org/abs/1606.09016v2.},  some relation between $2$-cells in $\s_{MLLu}$ and multiplicative proof structures with units  are stated. In particular, we achieve a cut-elimination rules correspondence, a one-to-one correspondence between proof structures and sets of equivalent $2$-cells modulo twisting relations and a cut-elimination result.
\vspace{-.4cm}
\subsection{Proof diagrams with control for $\MLLc$}

 We finally extend proof diagrams with control to the general case of $\MLLc$.
 

\begin{defi}
The \emph{control polygraph of multiplicative linear logic with constants} $\Mlltc$ is given by
\begin{multicols}{2}
\begin{itemize}
\item $\Mlltc_0=\{\pro \}$;
\item $\Mlltc_1=\fmllc \cup \{L= \twocell{L},R=\twocell{R}\}$;
\end{itemize}
\end{multicols}
\begin{itemize}
\item $\Mlltc_2=\begin{Bmatrix}
1: & \pro &\frr &L, 1,R  &=&\twocell{vc *1 (L *0 pit1 *0 R)}\\
\bot: & \pro &\frr &\bot  &=& \twocell{bot *1 pitbot} \\
\end{Bmatrix}\cup  \Mllt_2$
\item $ \Mlltc_3=  \Mlltc_{Twist}$  is made of rules in $\Mllt_{Twist}$ plus the following twisting relations:
\vspace{-.3 cm}
$$
\xymatrix{ \twocell{topA *1 (bot *0 1) *1  s *1( pitA *0 pitbot) } \ar@3[r] & \twocell{ topA *1 (1 *0 bot)  *1( pitA *0 pitbot) }},
\qquad
\xymatrix{ \twocell{topA *1 (1 *0 bot)*1 (s) *1 (pitbot *0 pitA)} \ar@3[r] & \twocell{ topA *1 (bot *0 1) *1 (pitbot *0 pitA) }};
$$\vspace{-.5 cm}
\end{itemize}

\end{defi}
					

\begin{teor}[Controlled proof diagram correspondence in $\Mlltc$]\label{corrMLLc}
$$\vdash_{MLLu} \Gamma \Leftrightarrow \exists \phi \in \Mlltc \mbox{ such that } \phi: \pro \frr L,\Gamma, R.$$
\begin{proof}
The proof can be given extending the one of Theorem \ref{corrMLL}.
To prove the left-to-right implication $\Rightarrow$ we should to consider the following two additional cases:
\begin{itemize}
\item If just one inference rule occurs $d(\Gamma)$, then it could also be a $1$,  $\Gamma= 1$ and $\phi_{d(\Gamma)}= 1: \pro \frr L, 1,R$;

\item If the last of the $n+1$ inference rules appearing in $d(\Gamma) $ is a $\bot$ (unary), $\Gamma= \Gamma', \bot$, then, by inductive  hypothesis, there is a  diagram $\phi_{\Gamma'}: \pro \frr L, \Gamma', R$ and $\phi_\Gamma= (L,\bot, \id_{\Gamma'},  R) \circ \phi_{\Gamma'}$;

\end{itemize}

In order to prove sequentialization, i.e. the right-to-left implication $\Leftarrow$, we have to consider the following two additional cases:
\begin{itemize}
\item If  $|\phi|_{ \Mlltc}=1$ then $\phi$ is an elementary diagram. The elementary diagrams with source $\pro$ and target $L,\Gamma, R$ with $\Gamma \in \fmllc^*$ are atomic made of a unique $2$-cell of type $Ax_A: \pro \fr L, A, A^\bot, R$ for some  $A\in \fmllc$ but also $1: 0 \fr L,1,R$. The associated sequent $\vdash 1$ is derivable in $\MLLc$;

\item Otherwise we should consider the case if there is  $2$-cell of type $\bot $. Then $\phi=(\id_{L, \Delta} , \bot, \id_{\Delta', R})\circ \phi'$ with  the diagram $\phi'$  sequentializable by hypothesis since $|\phi|_{\Mlltc}= |\phi' |_{\Mlltc} +1$
\end{itemize}
\vspace{-.6cm}
\end{proof}
\end{teor}

\vspace{-.6cm}


\section{Conclusion and future work}
\vspace{-.2cm}

We have presented \textit{proof diagrams}, a particular class of string diagrams suitable for interpreting linear logic proof derivations. In particular, such settings exhibit an internal correction criterion as we have shown a one-to-one correspondence between $MLL$, with or without units, (one-sided) sequent calculus proof derivations (with explicit exchange rules) and proof diagrams. Moreover, the sequentializability of a proof diagram, i.e. whether it corresponds to a proof in sequent calculus, depends on the number of inputs and outputs only, and can be verified in linear time.

Our results raise an important question about the quotient set over proofs introduced by proof diagrams, and how it relates to that performed by proof nets.

For this, let $\sim$ be the equivalence relation over proof derivations induced by proof diagrams equivalence $\simeq$ in $(\Mlltc)^*$. Then, one the one hand, $\sim$ captures all commutations of reversible inference rules $\parr$ and $\bot$ by the interchange rule and twisting relations. On the other hand, this is not the case for $\otimes$ and $Cut$: let $\alpha, \beta \in \{\otimes, Cut\}$, then $\sim$ equates only permutations of the kind that follows 
\vspace{-.3cm}
\begin{center}{\tiny \begin{tabular}
{p{4cm} p{1,5cm}  p{4cm}}
\begin{prooftree}
\AxiomC{$\overset  1 \vdots$}
\noLine
\UnaryInfC{$\vdash \s, A $}

\AxiomC{$\overset  2 \vdots$}
\noLine
\UnaryInfC{$\vdash B, \Gamma, C$}

\RightLabel{$ \alpha $}
\BinaryInfC{$\vdash \s, \alpha(A, B),  \Gamma , C$}
\AxiomC{$\overset  3 \vdots$}
\noLine
\UnaryInfC{$\vdash D, \Delta $}
\RightLabel{$ \beta $}

\BinaryInfC{$ \vdash \s , \alpha (A, B) , \Gamma, \beta(C, D), \Delta$}
\end{prooftree}

&
\begin{prooftree}
\AxiomC{~}
\noLine
\UnaryInfC{$~$}
\noLine
\UnaryInfC{$~$}
\noLine
\UnaryInfC{$~$}
\noLine
\UnaryInfC{$~$}
\noLine
\UnaryInfC{$~$}
\noLine
\UnaryInfC{$\sim$}
\end{prooftree}
&

\begin{prooftree}
\AxiomC{$\overset  1 \vdots$}
\noLine
\UnaryInfC{$\vdash \s, A $}

\AxiomC{$\overset  2 \vdots$}
\noLine
\UnaryInfC{$\vdash B, \Gamma, C$}

\AxiomC{$\overset  3 \vdots$}
\noLine
\UnaryInfC{$\vdash D, \Delta $}

\RightLabel{$ \beta $}
\BinaryInfC{$\vdash A , \Gamma,   \beta(C, D), \Delta$}
\RightLabel{$ \alpha $}
\BinaryInfC{$ \vdash \s , \alpha (A , B) ,\Gamma, \beta(C ,D) , \Delta$}

\end{prooftree}
\end{tabular},}\end{center}
that is, $\otimes$ or $Cut$ permutations that do not change the order of the leafs in a derivation tree. 

It follows that proof nets equivalence is coarser than proof diagrams one, proof nets equate more. For an actual example, consider the linear logic sequent $B\otimes C, A\otimes D$: this latter exhibits two different derivations that correspond to the following two non-equivalent proof diagrams 
$$\twocell{(dia1 *0 dia2 *0 dia3) *1 (L *0 topA *0 topB *0 R *0 L *0 topC *0 R *0 L *0 topD *0 R)*1 (L *0 1 *0 tenc *0 R *0 L *0 1 *0 R) *1 (L *0 s *0 R *0 L *0 1 *0 R) *1 (L *0 1 *0 tenc *0 R)} \not\simeq
\twocell{(dia1 *0 dia3 *0 dia2)*1 (L *0 topA *0 topB *0 R *0 L *0 topD *0 R*0 L *0 topC *0 R)*1 (L *0 s *0 R *0 L *0 1 *0 R *0 L *0 1 *0 R) *1 (L *0 1 *0 tenc *0 R *0 L *0 1 *0 R) *1 (L *0 s *0 R *0 L *0 1 *0 R) *1 (L *0 1 *0 tenc *0 R)*1 (L *0 s *0 R)} \;.$$ 
\vspace{-.1 cm}
On the other hand, the two proof derivations have the same proof net.

We conjecture that, in order to recover the whole proof equivalence induced by proof nets, we should extend control polygraph rewriting with the possibility to permute $Ax$ and $1$ gates' position in a diagram. Anyway, this is not related to our complexity result for sequentialization. Indeed, proof diagrams exhibit a local sequentialization criterion which is ruled out in proof nets by complexity arguments (P. Lincoln and T. Winkler \cite{MLLcNP}, W. Heijltjes \cite{noMLL}), due to the number of jumps to check. Crucial in our settings is the fact that $\bot$ gates have a specific position in diagrams, that one can interpret as a jump assignment: for example, given a $\bot$ gate, we can point its jump to the unique gate of type $Ax$ or $1$ connected to the left-nearest $L$ string. In particular, this means that equivalent proof diagrams in $(\Mlltc)^*$ may correspond to different jump assignments on the same proof net.

We believe this work suggests several future research directions. In particular, in the near future, we will focus on extending the present results to the multiplicative-exponential linear logic fragment.


%

\end{document}

%% file: 0_commands.tex
\usepackage{tikz}
\usepackage{amssymb}
\usepackage{verbatim} 
\usetikzlibrary{matrix}
\usepackage{color} 
\usepackage{stmaryrd}
\usepackage{amsfonts} 
\usepackage{amsmath} 
\usepackage{amsthm}
\usepackage[all,cmtip]{xy}
\usepackage{catex} 			
\usepackage{fancybox}
\usepackage{relsize}
\usepackage{cmll} 
\usepackage{bussproofs}
\usepackage{multicol}

\newtheorem{teor}{Theorem}

\newtheorem{oss}{Remark} 
\newtheorem{prop}{Proposition}

\theoremstyle{definition}
\newtheorem{defi}{Definition}

\usepackage[normalem]{ulem}

\newcommand{\nota}{\noindent\textbf{Notation. }}

\newcommand{\fr}{\rightarrow}						
\newcommand{\frr}{\Rightarrow}						

\newcommand{\R}{\mathcal R}

\newcommand{\N}{\mathbb N}
\newcommand{\s}{\Sigma}

\newcommand{\sig}{\mathcal S} 						
\newcommand{\pro}{\; \square \;}					

\newcommand{\D}{\phi}

\newcommand{\id}{\textbf{id}}

\newcommand{\cutelim}{\xymatrix@C-1pc{{} \ar@3[r]& {}}}
\newcommand{\vcutelim}{\xymatrix@R-0.5pc{~\ar@3[d] \\~ }}
\newcommand{\swap}{\twocell{s}}
\newcommand{\swapl}[2]{T_{#1 ,#2}}

\newcommand{\Mllt}{\tilde{\mathfrak{M}}}                
\newcommand{\Mlltc}{{\tilde{\mathfrak{U}}}}			

\newcommand{\fmll}{{\mathfrak{F}_{M\ell \ell}}}                
\newcommand{\fmllc}{{\mathfrak{F}_{M\ell \ell _u}}}                

\newcommand{\MLL}{\mathit{MLL}}				
\newcommand{\MLLc}{\mathit{MLL_u}}				

\deftwocell[crossing]{s : 2 -> 2}
\deftwocell[dots]{d : 2 -> 2}

\deftwocell[rectangle]{g : 2 -> 2} 				
\deftwocell[rectangle]{g35 :3 -> 5}				
\deftwocell[polygon]{g12:1 -> 2}				
\deftwocell[polygon]{g21 :2 -> 1}				
\deftwocell[text=\otimes,yellow]{net : 1 -> 2}			

\deftwocell[text=\alpha, white]{a : 2 -> 2} 			
\deftwocell[text=\phi, white]{phi : 2 -> 2} 			
\deftwocell[text=\phi', white]{phi1 : 2 -> 2} 			
\deftwocell[text=\psi, white]{psi : 2 -> 2} 			
\deftwocell[text=\sigma, white]{sigma : 2 -> 2} 		
\deftwocell[text=\sigma', white]{sigma1 : 2 -> 2} 		
\deftwocell[text=\chi_u, white]{up : 2 -> 6} 				
\deftwocell[text=\chi_d, white]{do : 6 -> 2} 				

\deftwocell[rectangle, white]{bigt : 8 -> 8} 			
\deftwocell[rectangle, white]{bigt12 : 12 -> 12} 				
\deftwocell[text=BT{(}W W' {)}, white]{btW : 8 -> 8} 			

\deftwocell[crossing2]{ln : 3 -> 3}				
\deftwocell[crossing1]{rn : 3 -> 3}				

\deftwocell[text=1, white]{dia1: 0 -> 2}
\deftwocell[text=2, white]{dia2: 0 -> 2}
\deftwocell[text=3 , white]{dia3: 0 -> 2}

\deftwocell[mid = \mathcal{C}]{backmC : 0 -> 0}				
\deftwocell[mid = \mathcal{D}]{backmD : 0 -> 0}				
\deftwocell[mid = \mathcal{E}]{backmE : 0 -> 0}				

\deftwocell[text=\circ, white]{circ : 2 -> 1} 			

\deftwocell[mid = A]{topA : 0 -> 1}						
\deftwocell[mid = {?}A]{topAw : 0 -> 1}					
\deftwocell[mid = A^\bot]{topAb : 0 -> 1}					
\deftwocell[mid = B^\bot]{topBb : 0 -> 1}					
\deftwocell[mid = B]{topB : 0 -> 1}						
\deftwocell[mid = C]{topC : 0 -> 1}						
\deftwocell[mid = D]{topD : 0 -> 1}						

\deftwocell[mid = \Gamma]{topGam : 0 -> 2}						
\deftwocell[mid = \Gamma']{topGam1 : 0 -> 2}						
\deftwocell[mid = {?}\Gamma]{topGamw : 0 -> 2}						
\deftwocell[mid = \sigma{(} {\Gamma } {)}]{topsiG : 0 -> 2}						
\deftwocell[mid = \Sigma]{topSig : 0 -> 2}						
\deftwocell[mid = \Delta]{topDel : 0 -> 2}						
\deftwocell[mid = L]{topL : 0 -> 1}						
\deftwocell[mid = R]{topR : 0 -> 1}						
\deftwocell[mid = W]{topW : 0 -> 2}						
\deftwocell[mid = W']{topW1 : 0 -> 2}						
\deftwocell[mid = W'']{topW2 : 0 -> 2}						

\deftwocell[mid = out {(}\alpha{)} ]{outa : 2 -> 2}						
\deftwocell[mid = in {(}\alpha{)} ]{ina : 2 -> 2}						

\deftwocell[mid = A]{midA : 1 -> 1}						
\deftwocell[mid = B]{midB : 1 -> 1}						

\deftwocell[mid = A]{pitA : 1 -> 0}						
\deftwocell[mid = {?}A]{pitAw : 1 -> 0}						
\deftwocell[mid = {!}A]{pitAo : 1 -> 0}						
\deftwocell[mid = A^\bot]{pitAb : 1 -> 0}					
\deftwocell[mid = B]{pitB : 1 -> 0}						
\deftwocell[mid = C]{pitC : 1 -> 0}						
\deftwocell[mid = \bot]{pitbot : 1 -> 0}						
\deftwocell[mid = 1]{pit1 : 1 -> 0}						
\deftwocell[mid = L]{pitL : 1 -> 0}						
\deftwocell[mid = R]{pitR : 1 -> 0}						

\deftwocell[mid = A \otimes {B}]{pitAtenB : 1 -> 0}						
\deftwocell[mid = A \parr {B}]{pitAparB : 1 -> 0}						
\deftwocell[mid = B \otimes {C}]{pitBtenC : 1 -> 0}						
\deftwocell[mid = B \parr {C}]{pitBparC : 1 -> 0}						
\deftwocell[mid = \Gamma]{pitGam : 2 -> 0}						
\deftwocell[mid = \Gamma']{pitGam1 : 2 -> 0}						
\deftwocell[mid = {?}\Gamma]{pitGamw : 2 -> 0}						
\deftwocell[mid = \Delta]{pitDel  : 2 -> 0}						
\deftwocell[mid = \Delta']{pitDel1  : 2 -> 0}						
\deftwocell[mid = \sigma{(} {\Gamma } {)}]{pitsiG : 2 -> 0}						
\deftwocell[mid = W]{pitW : 2 -> 0}						
\deftwocell[mid = W']{pitW1 : 2 -> 0}						
\deftwocell[mid = W'']{pitW2 : 2 -> 0}						

\deftwocell[text=\phi, white]{phi11 : 1 -> 1} 			
\deftwocell[mid = F]{topF : 0 -> 1}						
\deftwocell[mid = G]{topG : 0 -> 1}						
\deftwocell[mid = G]{pitG : 1 -> 0}						
\deftwocell[mid = G\circ {F}]{pitGcircF : 1 -> 0}						

\deftwocell[text=\otimes,yellow]{ten : 2 -> 1}		
\deftwocell[rectangle]{ax : 0 -> 2}				
\deftwocell[rectangle]{cut : 2 -> 0}				
\deftwocell[text=\parr, orange]{par : 2 -> 1}			
\deftwocell[circle, white]{v : 0 -> 1}				
\deftwocell[circle, black]{bot : 0 -> 1}					

\deftwocell[text=A,lightgray]{axA : 0 -> 2}				
\deftwocell[text=A^\bot,lightgray]{axAb : 0 -> 2}			

\deftwocell[text=A,lightgray]{cutA : 2 -> 0}				
\deftwocell[text=A^\bot,lightgray]{cutAb : 2 -> 0}			

\deftwocell[text=?,white]{D : 1 -> 1}				
\deftwocell[text=?,white]{C : 2 -> 1}				
\deftwocell[text=?,white]{W : 0 -> 1}				
\deftwocell[text=!,white]{P : 3 -> 3}				

\deftwocell[rectangle]{axt : 2 -> 4}			
\deftwocell[rectangle]{cutt : 4 -> 2}			

\deftwocell[text=\otimes,yellow]{tenc : 4 -> 1}		
\deftwocell[rectangle]{axc : 0 -> 4}			
\deftwocell[rectangle]{cutc : 4 -> 0}			
\deftwocell[rectangle,white]{vc : 0 -> 3}		

\deftwocell[text=A,lightgray]{axcA : 0 -> 4}					
\deftwocell[text=A^\bot,lightgray]{axcAb : 0 -> 4}				

\deftwocell[text=\otimes,yellow]{tenct : 5 -> 2}			
\deftwocell[rectangle]{cutct : 6 -> 2}			

\deftwocell[lefthalfcircle,red]{L : 1 -> 1}			
\deftwocell[righthalfcircle,blue]{R : 1 -> 1}			

\deftwocell[text=AX, lightgray]{elax : 2 -> 2} 			
\deftwocell[text=Cut, lightgray]{elcut : 2 -> 2} 			
\deftwocell[text=c,brown]{tenpar : 2 -> 2} 	
